\documentclass[aps,twocolumn,groupedaddress,showpacs]{revtex4}
\usepackage{amsmath,amscd,amssymb,epsfig}

\usepackage{amsmath}
\usepackage{graphicx}

\newcommand{\m}[1]	{\langle #1 \rangle}

\begin{document}
\bibliographystyle{apsrev}

\title[]{Origin of hyperdiffusion in generalized Brownian motion}
\author{P. Siegle}
\author{I. Goychuk}
\author{P. H\"anggi}
\affiliation{Institute of Physics, University of Augsburg,
Universit\"atsstr. 1, D-86135 Augsburg, Germany}

\date{\today}

\begin{abstract} 
We study a minimal non-Markovian 
model of superdiffusion which originates from long-range velocity correlations 
within the generalized Langevin equation (GLE)
approach. The model allows for a three-dimensional Markovian embedding. 
The emergence of a transient hyperdiffusion, 
$\langle \Delta x^2(t)\rangle \propto t^{2+\lambda}$, 
with $\lambda\sim 1-3$   is detected in tilted washboard 
potentials before it ends up in a ballistic asymptotic regime.
We relate this phenomenon to a transient heating of particles 
$T_{\rm kin}(t)\propto t^\lambda$ from the thermal bath temperature
$T$ to some maximal kinetic temperature 
$T_{\rm max}$. This hyperdiffusive transient regime ceases when the
particles arrive at the maximal kinetic temperature.

\end{abstract}


\maketitle

%\section{Introduction}
Anomalous diffusion is met across many different branches of physics, from
charge transport processes in amorphous materials to 
plasma physics and biophysics,
and there are many different theories reflecting a variety of
underlying physical mechanisms 
\cite{Scher,Shlesinger,Hughes,Bouchaud,Metzler,BarkaiSilbey,Friedrich}. 
One of fundamental approaches
is based on the generalized Langevin equation (GLE)
\cite{Kubo,Zwanzig,HTB90,Coffey,WeissBook,Wang,Morgado,Pottier,Bao,
Bao05,Kupferman,
Goychuk07,Deng,Goychuk,Siegle,Xia}, see Eq. (\ref{GLE})
below.
Remarkably, it can be derived from a Hamiltonian
dynamics of a particle that bilinearly couples 
to a thermal bath of harmonic oscillators 
characterized by the bath spectral density $J(\omega)$
\cite{Zwanzig,HTB90,WeissBook,Kupferman}.
This model is 
capable to describe all types of anomalous diffusion 
$\langle \Delta x^2(t)\rangle\sim 2 D_{\alpha}t^\alpha/\Gamma(1+\alpha)$ 
within a unified framework and the index $\alpha$ reflects just the 
low-frequency behavior
of the spectral bath density $J(\omega)\propto \omega^{\alpha}$ 
for $0<\alpha<2$
\cite{Grabert,WeissBook}. Namely, the case $0<\alpha<1$ 
corresponds to anomalously slow diffusion, or subdiffusion; 
the case $\alpha=1$ to normal diffusion, and the case 
$1<\alpha<2$ to superdiffusion. The ballistic diffusion, 
$\langle \Delta x^2(t)\rangle\sim D_{2}t^2$,
is attained for all spectral densities with $\alpha>2$ at low frequencies 
and the case of
$J(\omega)\propto \omega^2$ is marginally ballistic (a special case).
This finding is without a potential,
or under a constant force. Then the occurrence of hyperdiffusion,
$\langle \Delta x^2(t)\rangle \propto t^{\alpha}$
with $\alpha>2$, 
is not possible
if Brownian particles are 
initially thermalized \cite{remark}.  
Several circumstances are of special
interest. First, GLE diffusion is almost always ergodic, except for
the ballistic case \cite{Morgado,Bao05,Deng,Siegle} studied also below. 
Similarly, anomalous diffusion based on continuous time
random walks is weakly non-ergodic \cite{new1}. Second, for a constant 
force $F$, the diffusion
coefficient is proportional to the bath temperature, i.e. 
$D_{\alpha} \propto T$ and a generalized Einstein-Stokes relation 
holds \cite{WeissBook}.
For nonlinear forcing, e.g. in tilted washboard potentials, this kind
of anomalous diffusion is not sufficiently investigated and offers
surprises. In particular,  a hyperdiffusive regime occurs with 
$\alpha$ greatly enhanced to $\alpha_{\rm eff}\sim 3-5$ 
\cite{Lu,Siegle}. This puzzling nonlinear and nonequilibrium
effect is the focus of this study.

Such anomalous diffusion allows for Markovian embeddings of 
surprisingly small dimensions which suffice normally in practice 
\cite{Goychuk,Siegle}. A Markovian embedding is natural given 
that the underlying
Hamiltonian dynamics is Markovian. However, it formally has an infinite
dimension for a thermal bath considered in the thermodynamic limit. Surprisingly, 
the practical embedding dimension using some auxiliary 
stochastic variables can be quite small.

The purpose of this Letter is to give a  physical 
explanation of the observed hyperdiffusive anomaly as a transient
heating of particles with their kinetic temperature defined
via the velocity variance $\langle \Delta v^2(t)\rangle 
\propto T_{\rm kin}$ rising in accordance
to a transient power law, $T_{\rm kin}(t)\propto t^{\lambda}$, from the
bath temperature $T$  
to a maximal kinetic temperature $T_{\rm max}$
which depends on the duration of transient period through $F$, $T$, and
the amplitude of periodic potential $V_0$ ($T_{\rm kin}=T$, when $V_0=0$, or $F=0$). 
It can be very
large (thousands of $T$). Such a nonlinear heating mechanism
in fixed \textit{not alternating in time} 
applied fields is quite unusual, and, paradoxically, smaller bias strengths
$F$ yield
higher $T_{\rm max}$ values.

We start from the traditional GLE model in one selected direction 
for a particle of mass $m$ in the potential $V(x)$ subjected to
a linear friction with memory kernel $\eta(t)=(2/\pi)\int_0^{\infty}d\omega
J(\omega)\cos(\omega t)/\omega$ and random force 
$\xi(t)$ of zero-mean:
\begin{eqnarray}\label{GLE}
  m\ddot x+\int_0^t\eta(t-t')\dot x(t')dt'+
  \frac{\partial}{\partial x}V(x)=\xi(t)\;.
\end{eqnarray}
The random force is Gaussian and fully characterized by its 
autocorrelation function satisfying the fluctuation-dissipation relation
\begin{equation}\label{FDR}
  \m{\xi(t)\xi(t')}=k_BT\eta(|t-t'|)
\end{equation}
which in turn is a consequence of the fluctuation-dissipation theorem.

A necessary condition for the emergence of  superdiffusion asymptotically 
($t\!\to\!\infty$) within the considered
class of models is zero integral friction \cite{Morgado,Siegle}, i.e.
$\lim_{t\to\infty}\int_0^t\eta(t')dt'=0$. 
The memory kernel thus must be positive 
at times $t'=t$, yielding $\eta(0)>0$, cf. Eq. (\ref{FDR}), and possess 
a negative part. The simplest model which satisfies these two conditions
is \cite{Bao05}:
\begin{equation}\label{kernel}
  \eta(t)=\eta\left[2\delta(t)-\nu e^{-\nu t}\right]\;.
\end{equation}
It will be considered in the following and corresponds to a spectral
bath density $J(\omega)$ which is cubic for $\omega\ll \nu$ (typifying
e.g. acoustic bulk phonons in solids \cite{WeissBook}) and linear for 
$\omega\gg \nu$. The corresponding spectral power of the noise 
$S(\omega)$ corresponds to the white noise (for $\omega\gg \nu$),
i.e. $S(\omega)=const$,
with the small-frequency part of the spectrum smoothly cut, 
so that $S(\omega)\propto \omega^2$ for $\omega\ll \nu$.

Furthermore, the autocorrelation function (ACF) of velocity fluctuations, 
$\Delta v(t)=v(t)-\langle v(t)\rangle$, in the 
absence of deterministic force, or
under a constant forcing obeys for this minimal model
\begin{eqnarray}
\langle \Delta v(t)\Delta v(t')\rangle & = & v_T^2 
\Big \{ \frac{\nu}{\nu+\gamma}\nonumber \\
 & &+ \frac{\gamma}{\nu+\gamma}
\exp[-(\nu+\gamma)|t-t'|] \Big \}
\end{eqnarray}
with $\gamma=\eta/m$, provided that the velocities are 
initially thermally distributed with 
$\sqrt{\langle \Delta v^2\rangle}=v_T=\sqrt{k_BT/m}$. 
This follows from the Laplace-transformed result for this 
quantity for arbitrary kernels 
\cite{Kubo}: 
$\tilde K_v(s)=v_T^2/[s+\tilde \eta(s)/m]$. 
In other words, the ACF
exponentially decays, but to a non-zero constant, which 
is the reason for nonergodicity
and ballistic diffusion. 
Clearly, this is the simplest of possible models, yet physically
reasonable. 
The model is non-Markovian,
but it allows for a three-dimensional Markovian embedding \cite{Bao05}:
\begin{eqnarray}\label{embedding}
  \dot x(t)&= &v(t) \nonumber \\
  m\dot v(t)&=& -\frac{\partial}{\partial x}V(x)-u(t)-\eta v(t)+
  \sqrt{2k_BT\eta}\zeta(t) \nonumber \\
  \dot u(t)&=&-\nu\eta v(t)-\nu u(t)+\nu
  \sqrt{2k_BT\eta}\zeta(t)\;,
\end{eqnarray}
where $\zeta(t)$ is a zero-centered white Gaussian noise 
with autocorrelation function $\m{\zeta(t)\zeta(t')}=\delta(t-t')$.
By integrating out the  auxiliary variable $u(t)$ in the above
equations (projecting onto the $x-v$ plane), 
it is not difficult to show that this leads to the GLE in Eqs. (\ref{GLE}, 
\ref{FDR}) with the memory kernel of Eq. (\ref{kernel}), if the
initial value of $u(0)$ is Gaussian distributed with zero mean
and variance $\langle u^2(0)\rangle =k_B T\eta $.
This is assumed in the following. The initial velocities are 
thermally distributed.

\begin{figure}
\includegraphics [width=\linewidth]{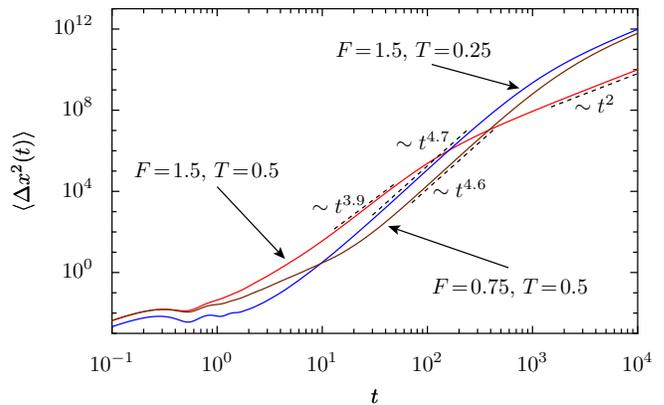}
\caption{(Color online)
Biased diffusion in a washboard potential with strength $V_0=1$
for different values of the bias $F$ and the bath temperature $T$.
The mean squared displacement exhibits transient hyperdiffusion
before it reaches the asymptotically ballistic diffusion regime.
The transient hyperdiffusion is enhanced upon decreasing $F$ and $T$.
The velocities are initially thermally distributed,
$\nu=0.25$, and $n=10^{4}$ trajectories are used for the ensemble averaging.
}
\label{Fig1}
\end{figure}

We consider the case of a washboard potential with period $x_0$ and amplitude
strength $V_0$ biased by a constant force $F$,
$V(x)=-V_0\cos(2\pi x/x_0)-Fx$.
It is convenient to transform Eq. (\ref{embedding}) into dimensionless quantities by scaling
time in units of $\gamma^{-1}$,
distance in $x_0$,
energy in $m (x_0\gamma)^2$ (which applies for $V_0$, $k_BT$, and $Fx_0$),
$u$ in $mx_0\gamma^2$, and $\nu$ in $\gamma$.
We have integrated the system 
in the corresponding nondimensional variables using a standard Euler algorithm 
with time step 
$\Delta t=10^{-4}$ for $\nu=0.25$, $V_0=1$, and varying $T$ and $F$. 
The behavior of the position variance for $10^4$ particles which started
at the origin with the velocities thermally distributed is depicted in Fig. \ref{Fig1}.
 A striking feature is the regime of
intermediate hyperdiffusion which ends in the ballistic regime.
This puzzling behavior, which seems to be rather general, was
not, however, explained before in physical terms. 

For this we notice that the diffusive
behavior can be related to the twice-integrated velocity autocorrelation function (ACF) 
$\langle \Delta v(t)\Delta v(t')\rangle$. 
In the stationary limit, when the ACF depends
on the difference of time arguments, the normal diffusive behavior emerges when the
integral of  ACF is finite. Its value defines the diffusion coefficient, which is 
temperature-dependent. For the thermally
distributed velocities and normal diffusion (singular limit for $\nu=0$) 
the ACF decays exponentially to zero from
$\langle \Delta v^2\rangle =k_BT/m$, with the decay rate $\gamma$. 
This yields 
the Einstein relation, $D(T)=k_BT/(m\gamma)$. For the anomalous GLE,
diffusion is described
by a spectral bath density $J(\omega)\propto \gamma_{\alpha}\omega^\alpha$ 
($0<\alpha <2$) 
which corresponds
to $\tilde \eta(s)=\eta_{\alpha}s^{\alpha-1}$ and yields
$\langle \Delta x^2(t)\rangle \sim 2 D_{\alpha}(T) t^{\alpha}/\Gamma(1+\alpha)$
asymptotically, 
the integral of the ACF for $\alpha\neq 1$ either diverges (superdiffusion), or it tends to zero 
(either subdiffusion, or bounded
motion in trapping potentials).
However, in the absence of forcing, or under a constant force a generalized Einstein
relation always holds, $D_{\alpha}(T)=k_BT/(m\gamma_{\alpha})$. 
Furthermore, it is easy to show
(from the exact expression for $\tilde K_v(s)$ for $V_0=0$) 
that for $\alpha >2$, the diffusion is asymptotically always ballistic.
This is why the occurrence of long-lasting 
hyperdiffusion is rather surprising. 
For the considered model we have  
$\langle \Delta x^2(t)\rangle \sim D_2(T)t^2$ asymptotically with
$D_2(T)=k_BT/m^*$, where $m^*=m(1+\gamma/\nu)$ is an effective mass. 
This asymptotics holds in the absence of a periodic potential, i.e. $V_0=0$.

\begin{figure}
\includegraphics [width=\linewidth]{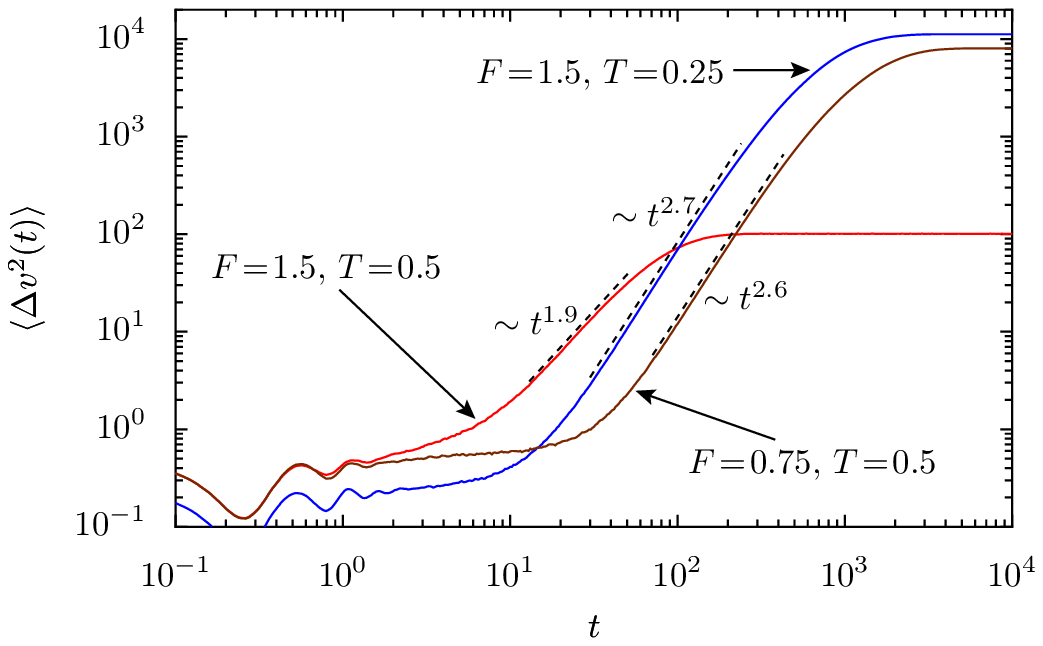}
\caption{(Color online)
The mean square fluctuation of the particles velocity for the biased
diffusion in Fig. \ref{Fig1} exhibits a  transient behavior,
in which the  kinetic temperature $T_{\rm kin}$ (see text)
grows from bath temperature to a maximal value $T_{\rm max}$.
The transient power law growth, $T_{\rm kin}(t)\propto t^{\lambda}$, 
explains the hyperdiffusive  regime with the corresponding exponent $2+\lambda$ 
in Fig. \ref{Fig1}.
Lower bath temperature and weaker biasing field lead surprisingly to 
stronger heating (see Fig. \ref{Fig4}).
}
\label{Fig2}
\end{figure}

\begin{figure}
\includegraphics [width=\linewidth]{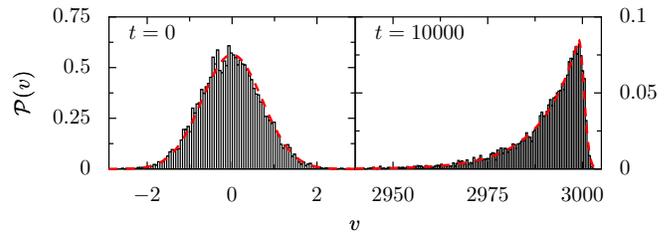}
\caption{(Color online)
Two snapshots of the velocity distribution $P(v,t)$, initial and at the end point of simulations
in Figs. \ref{Fig1}-\ref{Fig2} for $F=1.5$ and $T=0.5$.
$P(v,t)$ becomes shape invariant already for $t> 500$, after the kinetic temperature
reaches its maximum, see in Fig. \ref{Fig2}. After this happens, only the maximum of the distribution
 moves accelerating in time. 
The initial distribution is Gaussian with $\langle \Delta v^2(0)\rangle=T$.
The final distribution is strongly skewed:
its left slope ($v<2999$) is well fitted by an exponential, 
while the right slope ($v>2999$) remains
approximately Gaussian.
}
\label{Fig3}
\end{figure}

\begin{figure}
\includegraphics [width=1.0\linewidth]{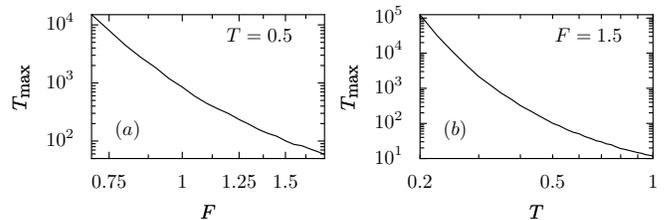}
\caption{Dependence of maximal kinetic temperature 
$T_{\rm max}$: (a) on applied force $F$ 
for a fixed thermal bath temperature $T$, 
and (b) on $T$ at a fixed $F$. 
}
\label{Fig4}
\end{figure}

If one switches on the periodic potential, the ballistic diffusion 
turns over into 
normal diffusion when $F=0$ \cite{Siegle}. Upon application of a 
constant force $F>0$, 
the particles will start to gradually  accelerate when they 
leave the attraction
domains of potential wells in the phase space due to the random kicks 
given by thermal noise. 
Their initial Maxwellian velocity distribution hence will not
hold forever. 
Gradually accelerating, the running particles 
undergo drastic transient heating above the bath temperature
due to multiple scattering on the 
periodic potential. Slower particles, however,  are
more strongly scattered backwards (i.e. are 
decelerated) by the potential wells in comparison with faster
ones. This in turn yields a growing width (see Fig. \ref{Fig2}) 
of the velocity distribution,
which becomes also skewed towards slower particles,  see Fig. \ref{Fig3}.
Indeed, the averaged kinetic energy per particle, $K=m\langle v^2\rangle/2$, 
can be decomposed as $K=K_m+K_T$. $K_m=m\langle v\rangle^2/2$
is the kinetic energy associated with mean velocity which asymptotically
is well described by $\langle v(t)\rangle=Ft/m^*$ when
the influence of periodic potential becomes negligible. 
The part 
$K_T=m\langle \Delta v^2\rangle/2\Doteq k_BT_{\rm kin}/2$ is
used to define the effective kinetic temperature $T_{\rm kin}(t)$.
Importantly, the action of $F$ results not only in the 
growing mean $\langle v(t)\rangle$, but also
generates a growing variance
of the velocity distribution, see in Fig. \ref{Fig2}.
It is a common practice to characterize different sorts of particles
with different kinetic temperatures e.g. in plasma physics 
\cite{volume10}. 
In a similar spirit,
we use a ``kinetic temperature'' notion 
which should be used with care as
it does not correspond to a thermodynamic temperature, but rather
simply characterizes the width of a nonequilibrium  velocity
distribution, see in Fig. \ref{Fig3}. 
Nevertheless,  it is a useful concept because it reflects a
important statistical aspect of the temperature; namely, 
that  temperature characterizes the width of the kinetic 
energy distribution. Even if
the relative spread of the velocity distribution around  the 
mean value is rather small at the end point of simulations (just a
few percents),  the mean kinetic energy is large and therefore 
the kinetic temperature can also be large, cf. Fig. \ref{Fig4}.

In our numerical experiments, the velocity distribution becomes
broadened and strongly skewed with the ``retarded'' tail of distribution
described by an exponential, rather than by a Gauss-Maxwell
distribution,  cf. in Fig. \ref{Fig3}. 
The variance of the distribution grows in time in
accordance with a power law, $T_{\rm kin}(t)\propto t^{\lambda}$, cf.
Fig. \ref{Fig2}. This explains the emergence of hyperdiffusion
$\langle \Delta x^2(t)\rangle \sim D_2(T_{\rm kin}) 
t^2=k_BT_{\rm kin}t^2/m^*
\sim t^{2+\lambda}$ in Fig. \ref{Fig1}. This regime is,
however, only transient. 
The duration of the transient period depends strongly on
the potential amplitude $V_0$,  the strength of the bias force $F$, and the bath
temperature $T$. 

The hyperdiffusive regime turns over into the ballistic diffusion
regime when the particles arrive at the maximal kinetic temperature
$T_{\rm max}$, compare Fig. \ref{Fig1} and Fig. \ref{Fig2}. This
nonlinear heating mechanism is quite unusual. The heating of plasmas
by time-varying stochastic fields is well-known. One of the pertinent
nonlinear mechanisms is the so-called Fermi acceleration \cite{Zaslavski}. 
It requires
but a time-varying driving field (or stochastically oscillating
boundary).   In our case, the ``heating'' field is however constant, and,
strikingly enough, the use of weaker bias fields heats up the particles
ever more strongly, see in Fig. \ref{Fig4}(a). However, much longer 
times are required then.  
Moreover, the
smaller the bath temperature is, the higher is the final kinetic
temperature, cf. Fig. \ref{Fig4}(b). Both effects are 
due to the fact that the transient
time scale becomes longer because the particles take on the kinetic
energy more slowly in the accelerating field $F$. 
The corresponding dependencies are stronger than exponential.
 Such a strong  sensitivity is surprising.
The qualitative physical explanation is as follows: 
The heating is caused by retardation of  flying particles when they pass over 
the trapping domains  while moving in the bias direction. 
It is
appreciably  strong as long as the averaged energy of the particles does
not substantially exceed the potential barrier height.
For smaller temperatures and
smaller bias forces it takes an exponentially larger time for the
particles to escape out of potential wells and to arrive on average at a 
sufficiently high kinetic energy,
so that their back scattering or 
deceleration cease to play a role. 
This leads to a
larger broadening of the velocity distribution \cite{remark2}.
The giant enhancement of ballistic diffusion reminds one of
giant enhancement of normal diffusion \cite{Reimann}. However, the underlying physical
mechanisms are quite different.

The physical systems where the discussed hyperdiffusive heating effect might
be relevant are  dusty plasmas where  heavy tracer 
particles collide with the 
gas of light particles serving as a thermal bath. 
Even if our GLE description in this case is not
directly applicable, there exists some partial
correspondence between the superdiffusive
fractional GLE results and the fractional Kramers equation by
Barkai and Silbey \cite{BarkaiSilbey}.
Such a correspondence
is surprising because  both descriptions
are different, see e.g. in \cite{Coffey}. The latter
scheme derives from linear Boltzmann equations with a fractional
scattering integral accounting for scattering events which are 
power-law distributed in time. 
Our physical explanation of the transient 
heating mechanism is more generally applicable, i.e. it is not
restricted by the present GLE model. In particular, it is expected to work  
for the fractional 
kinetic equations like the fractional Kramers
equation, or a more
general one introduced recently by Friedrich {\it et al.} \cite{Friedrich}.
We are confident that our work will stimulate further studies and even an experimental 
validation of this intriguing hyperdiffusive behavior.

This work was supported by the German Excellence
Initiative via the Nanosystems Initiative Munich (NIM).

\newpage 

 \begin{figure}
\includegraphics [width=1\linewidth]{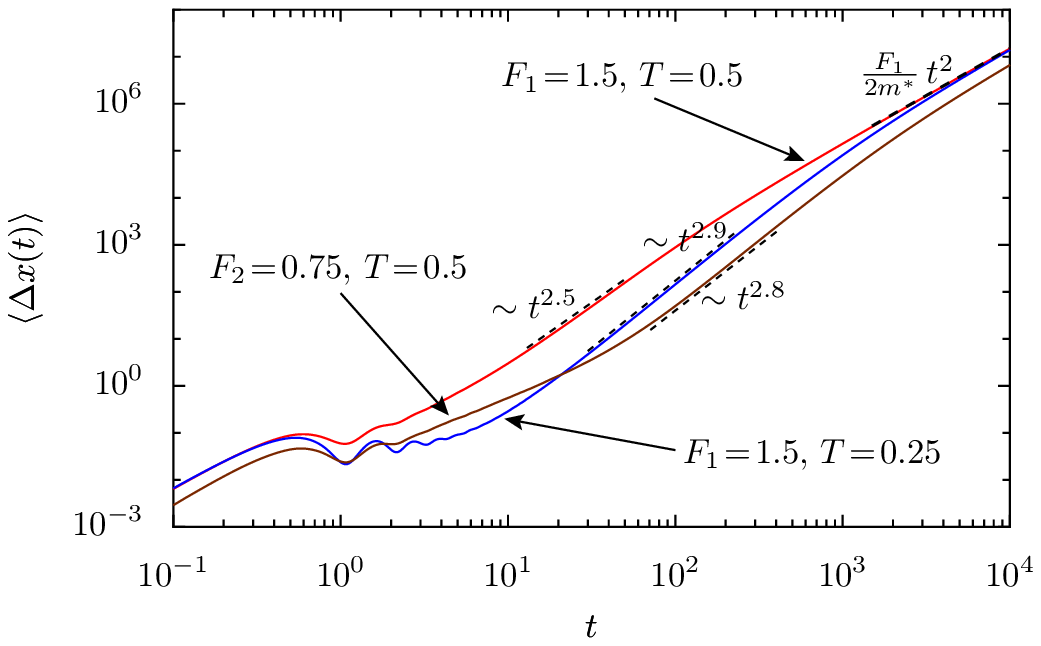}
%\label{Fig1}
FIG 1.: 
%\caption{
Dependence of mean displacement 
$\langle \Delta  x(t)\rangle$ on time. \\ \hspace{-1.4cm} 
The parameters are same as in Fig. 1 of main text. %}
\end{figure}

 \begin{figure}
\includegraphics [width=1\linewidth]{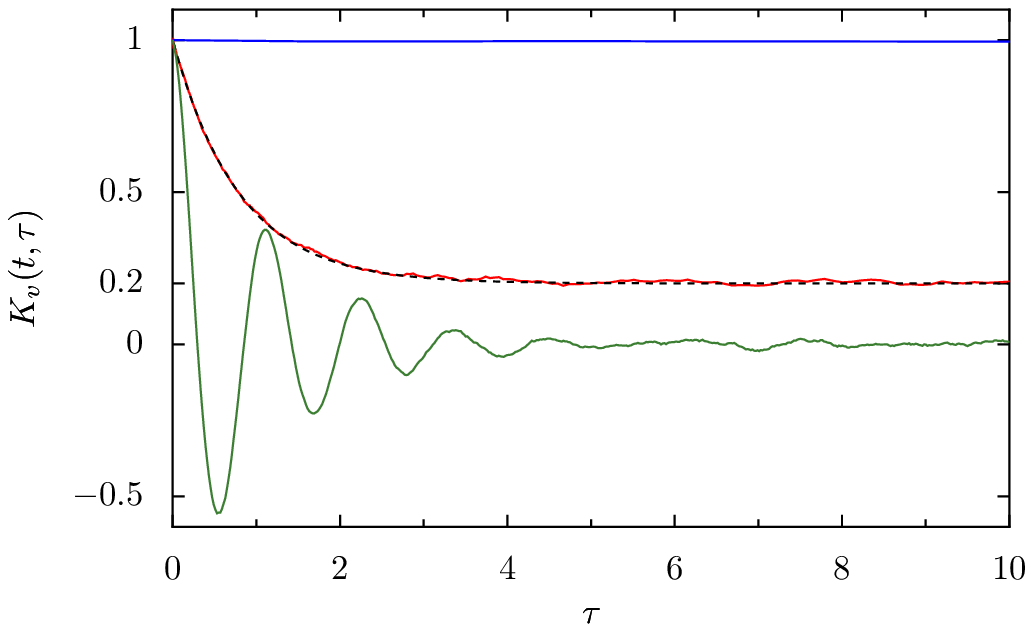}
%\label{Fig2}
FIG 2.: 
%\caption{
The normalized velocity autocorrelation function as a function of 
time lag $\tau$, in the absence
and in the presence of periodic potential for biased 
and unbiased diffusion. See \\ \hspace{-5.8cm}
 text for explanation. %}
\end{figure}

\begin{center}

\textbf{\large Supplement}

\end{center}

 Fig. 1 of this Supplementary Material depicts the dependence of the
mean particle displacement $\langle \Delta  x(t)\rangle$ on time $t$.
First, a regime of transient hyperacceleration is clearly seen which
accompanies the regime of transient hyperdifussion. Second,
asymptotically $\langle \Delta  x(t)\rangle$ is well described by
$\langle \Delta  x(t)\rangle\sim Ft^2/(2m^*)$ with the effective mass
$m^*=m(1+\gamma/\nu)$ which is $m^*=5 m$ for the studied parameters.

Fig. 2 depicts the normalized autocorrelation function 
$K_v(t,\tau)=\langle \Delta v(t+\tau)\Delta v(t)\rangle/\langle \Delta
v^2(t)\rangle$ of the velocity fluctuation, $\Delta v(t)=v(t)-\langle
v(t)\rangle$, as a function of the time lag $\tau$ for $t=500$ for
three cases: without periodic potential for a bias $F=1.5$ (red
line);  in the presence of periodic potential, $V_0=1$, for the same
$F=1.5$ (blue line), and for the same periodic potential in the
absence of bias, $F=0$.  In all cases $T=0.5$. The analytical result
in Eq. (4) of the main text (broken line) agrees well with numerics
(red line). The velocity ACF decays exponentially  to the constant
value $\nu/(\nu+\gamma)=0.2$  which indicates nonergodicity, as
non-zero correlations hold forever, for an arbitrary long time lag. 
Upon switching on the periodic potential, the ACF fails to decay (blue
line, the maximal kinetic temperature $T_{\max}$ is already reached
for studied parameters and $t\geq t_0=500$). This implies that
nonergodicity becomes even more pronounced. On the contrary, if one
switches off the bias $F\to 0$, the diffusion becomes normal and
ergodicity is restored in the periodic potential. In this case, the
ACF decays oscillatory to zero (green line in Fig. 2). Its integral is
finite and yields the normal diffusion coefficient (scaled in
$k_BT/m$).

Next let us discuss shortly the energetics of biased ballistic
superdiffusion. After the hyperdiffusion regime is over, the particles
have arrived at a maximal kinetic temperature and maximal broadening
of the kinetic energy distribution. The influence of the periodic
potential is then negligible.  The particles continue, however, to
accelerate in the applied field $F$, moving as quasi-particles  with
effective mass $m^*=m(1+\gamma/\nu)$, see in Fig. 1 of this
Supplementary Material.  Asymptotically the mean particle position
grows like $\langle x(t)\rangle \sim F t^2/(2m^*)$ and the work spent
by $F$ is therefore estimated as  $W \approx F\langle x\rangle \approx
F^2 t^2/(2m^*)$. Its action is: (i) to accelerate the (real) particles
to $\langle v\rangle$ with kinetic energy $K_m=m\langle
v\rangle^2/2=(m/m^*)W$, (ii) to heat them up to  $K_T=m\langle \Delta
v^2\rangle/2=k_B T_{\rm kin}(t)/2$. The rest is transferred as an
excess heat $Q$ to the thermal bath: $W=K_m+K_T+Q$. Asymptotically
$K_T\ll W,K_m,Q$ and therefore it can be neglected in the energy
balance. It thus follows that approximately,  $Q\approx W-K_m\approx
(m^*-m)W/m^*=\gamma W/(\gamma+\nu)$. For $\nu\ll \gamma$, most of the
acquired work is dissipated as excess heat. This is a
counter-intuitive conclusion, given that the integral friction is
zero. For the studied parameters, we find that $Q\approx 0.8\;W$.

\end{document}